\documentclass[twocolumn,aps]{revtex4-1}
\usepackage[dvipdfm]{hyperref}
\usepackage{graphicx}
\usepackage{amsmath}
\usepackage{hypernat}
\begin{document}
\title{Robust Dynamical Decoupling for Arbitrary Quantum States  of a Single NV Center in Diamond}

\author{J. H. Shim}
\author{I. Niemeyer}
\author{J. Zhang}
\author{D. Suter}
\affiliation{
 Fakult\"{a}t Physik, Technische Universit\"{a}t Dortmund, D-44221 Dortmund, Germany\\
}

\begin{abstract}
Dynamical decoupling is a powerful  technique for extending the coherence time (T$_2$) of  qubits.
We apply this technique to the electron spin qubit of a single nitrogen-vacancy center in type IIa diamond.
In a crystal with natural abundance of $^{13}$C nuclear spins,
we extend the decoherence time up to 2.2 ms.
This is close to the T$_1$ value of this NV center (4 ms).
Since dynamical decoupling must perform well for arbitrary initial conditions,
we measured the dependence on the initial state and compared the performance of different sequences
with respect to initial state dependence and robustness to experimental imperfections.\end{abstract}


\maketitle

\section{Introduction}

Sufficiently long coherence times (T$_2$) are amongst the most important criteria
for the realization of scalable quantum information processors.\cite{Ladd2010}
In solid state systems, it is often possible to extend the coherence time by removing
sources of decoherence from the host material, e.g. by isotopic engineering.\cite{Tyryshkin2011, Balasub2009}
If this is not possible, or to further reduce the detrimental effects of a noisy environment,
it is possible to use a series of control pulses applied to the qubits.
This approach, termed Dynamical Decoupling (DD),  refocuses the interaction between
system and environment by applying a stroboscopic sequence of inversion pulses to the qubits.
It has been tested on different systems with environments consisting of electronic or nuclear spin baths.
\cite{Du2009, Biercuk2009, Bylander2011, Lange2010, BarGill2012, alvarez2010}
The technique allows, e.g., to extend the decoherence time T$_2$ or to increase the sensitivity
of magnetic field sensors to ac magnetic fields.\cite{Lange2011, BarGill2012}

The usefulness of this approach for practical applications depends on the robustness of the performance
with respect to experimental imperfections, such as finite precision of control field amplitudes.
Since the number of control operations required for effective DD can be very large,
it is essential that the errors from the individual control operations do not accumulate.
This goal can be achieved by using fault-tolerant sequences, which are designed in such a way
that the errors of any individual control operation is cancelled by all the other operations
of a cycle.\cite{ryan2010, cai2011, souza2011}

Here, we test this approach on a single Nitrogen-Vacancy color center (NV center) in diamond.
The effectiveness of the DD technique has been tested with different types of diamond;
de Lange, et al.\cite{Lange2010} used a diamond with a relatively high concentration
of nitrogen impurities, which generates a strong electron-spin bath.
The other studies employed CVD grown diamond crystals with reduced native nitrogen concentrations,
in which the nuclear spin bath of the $^{13}$C nuclear spins is the major source of
decoherence.\cite{ryan2010, Naydenov2011, BarGill2012}
Here, we extend these earlier studies by testing the limits of DD in two respects:
first, we use DD to extend the dephasing time $T_2$ to nearly the value of the spin-relaxation time $T_1$.
To reach this limit, we have to apply hundreds of refocusing pulses.
With such a large number of pulses, the unavoidable imperfections of the individual gate operations
would normally destroy the coherence of the qubits.
To avoid this, we use robust DD sequences, which were designed such that the errors due to
the individual gate operations do not add up but cancel each other over the complete DD cycle.

When the effectiveness of DD sequences is demonstrated, it is often tempting to consider its
performance for a specific initial condition that is left unperturbed by the combination of DD and
the environmental perturbation.
However, in actual quantum computers, the state of the system is generally not known.
Accordingly, any serious test of DD performance must consider the performance of arbitrary
initial conditions.
We therefore present here a study in which we compare the performance of different
DD sequences over oll possible input states.

\begin{figure}
\includegraphics[width=\columnwidth]{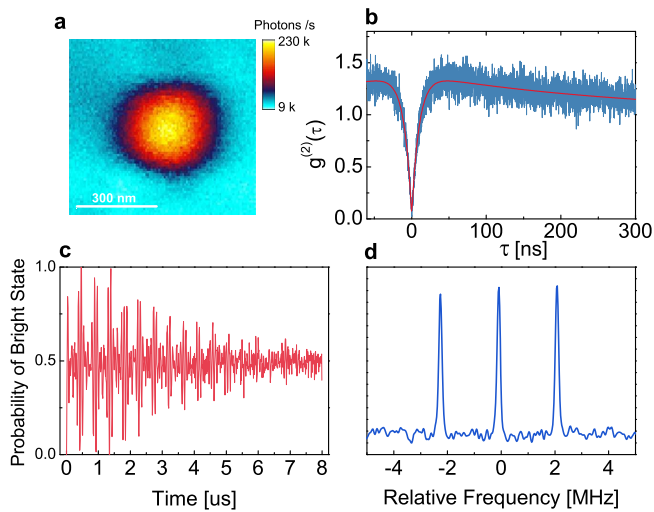}
\caption{(color online)(a) Confocal image (0.8 $\mu$m*0.8 $\mu$m) of single NV center (b) Second order correlation measurement by photon antibunching experiment. (c) Ramsey fringe (d) Fourier transformation of Ramsey fringe curve shows only three peaks due to hyperfine interaction with $^{14}$N in the NV center.  }\label{fig1}
\end{figure}

\section{Experimental setup}

In the experiments described below, we use a home-built confocal microscope
for optical addressing and detection of single NV centers.
Fig.~\ref{fig1} (a) shows, as an example, the scan image of one NV center.

A diode-pumped solid state laser working at a wavelength of  532 nm provides the optical excitation.
The acousto-optical modulator used for pulse shaping of the laser
provides an  extinction ratio of more than 57 dB.
To verify that we detect signals from single NV centers, we use a standard photon-antibunching
configuration with single photon detectors in combination with a fast time-correlation card.
Fig.~\ref{fig1} (b) shows a typical correlation signal trace.

We generate the microwave pulses with an arbitrary waveform generator.
Working at 2 GS/s, it generates the pulses around a carrier frequency of 250 MHz.
These pulses are upconverted to the target frequency in the range of 2.6 - 3.2 GHz
by mixing them with a cw  signal from a stable microwave source.
Unwanted sidebands are  filtered out by properly selected band-pass filters.
The amplified pulses pass through a 20 $\mu$m diameter copper wire
attached to the surface of the diamond crystal and terminated by 50 $\Omega$.
The leakage of microwave power at pulse-off time is $<-60$ dB.
Under typical experimental conditions, the duration of a 180 degree pulse is 40 ns.
A permanent magnet is used to apply a static magnetic field,
which is roughly aligned with the direction of the principal axis of the NV center.
A type IIa diamond crystal with a N concentration $<5$ ppb is the host material of the investigated NV centers.
All the measured curves in the present work are normalized in order to indicate the probability ($p$) of bright state ($m_S=0$).
As a reference for the normalization, maximum and minimum values from Rabi oscillation curve were taken for $p=1$ and $p=0$ respectively.

\section{Experimental results}

\subsection{Spin system}

In the natural abundance diamond crystal with reduced N concentration,
the $^{13}$C nuclear spin form a spin bath that induces decoherence on the NV center.\cite{Balasub2009, Mizuochi2009}
In the case of Hahn echoes, the deocherence due to the $^{13}$C nuclear spins is reduced significantly
when the free precession period between the excitation and refocusing pulse is a multiple of the nuclear spin
Larmor precession period.\cite{maze2008}
For the present study, we selected an NV center that has no  $^{13}$C nuclear spins in its immediate vicinity,
so the coupling to the nuclear spin bath is relatively weak and the decoherence correspondingly slow.
Figure \ref{fig1} (c) shows the Ramsey fringe signal measured from this center and Figure \ref{fig1} (d)
its Fourier transform.
We applied a magnetic field of 6.8 mT along the direction of the symmetry axis and excited the transition
at 2.68 GHz with resonant microwave pulses.

\subsection{Dynamical decoupling}

\begin{figure}
\includegraphics[width=\columnwidth]{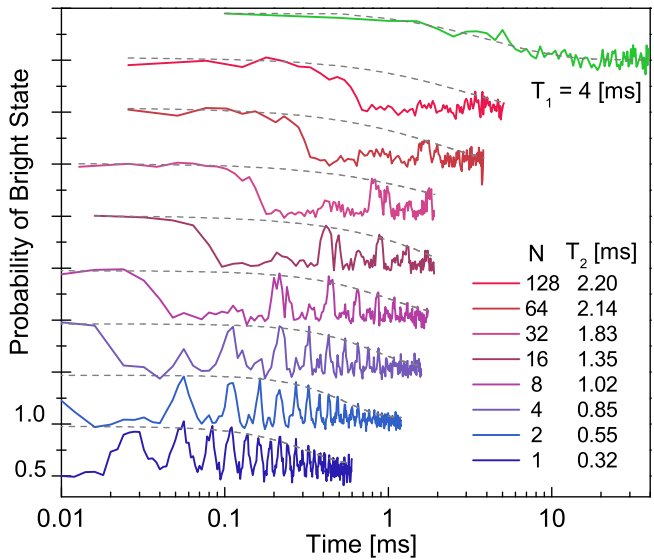}
\caption{(color online) Coherence decay curves for different dynamical decoupling sequences.
The bottom curve shows the result from the Hahn echo, the higher curves correspond to the CPMG sequence
with increasing number of refocusing pulses.
The gray dashed lines represent fits to the function  $\exp(-(\frac{t}{T_2})^n)$.
The initial state was chosen along the rotation axis of the refocusing pulses. The top curve shows T$_1$ relaxation of bright state.(T$_1$ = 4 ms)
}\label{fig2}
\end{figure}

The oldest and simplest pulse sequence for dynamical decoupling is the CPMG sequence -
a train of equidistant $\pi$-pulses around the same axis.
In Fig.~\ref{fig2}, we compare the signal decays for different CPMG sequences, starting with the Hahn echo
at the bottom and experiments with increasing numbers of refocusing pulses towards the top.
In each curve, we observe an initial fast decay, which is followed by a series of revivals.
The revivals are separated by $2 N \tau_L$, where $N$ is the number of $\pi$ pulses in the sequence
and $\tau_L \approx 73 \, \mathrm{\mu s}$ is the Larmor period of the $^{13}$C nuclear spins.
This implies that the refocusing is most effective when the nuclear spins constituting the environment
have completed an integral number of Larmor precessions between the pulses.
The dashed curves in Fig.~\ref{fig2} represent functions $\exp(-(\frac{t}{T_2})^n)$
with decay times $T_2$ and exponents $n$ fitted to the maxima of the Larmor revivals in the experimental data.
The decoherence times $T_2$ are also noted in the figure legend.

\begin{figure}
\includegraphics[width=\columnwidth]{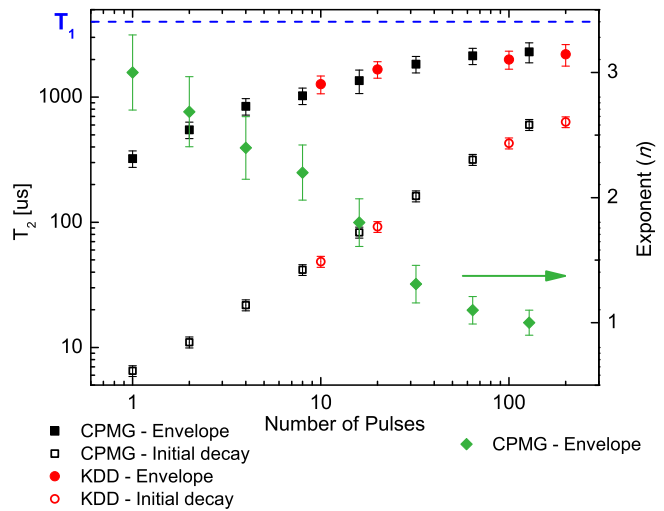}
\caption{(color online) T$_2$ and T$_2^0$ as a function of the number of pulses in CPMG and KDD sequences.
Black symbols are CPMG and red KDD.
Solid symbols represent T$_2$s obtained from the envelope of the revivals,
and open symbols are T$_2^0$s from the initial decay.
Green symbols represent the exponent $n$ in the decay function $\exp(-(\frac{t}{T_2})^n)$.
The initial state was chosen along the rotation axis of the CPMG pulses.
}
\label{fig3}
\end{figure}

Figure \ref{fig3} summarizes the increase of the decoherence time with the number of refocusing pulses.
The decoherence times from the CPMG experiment are shown as black squares, with filled squares
representing the values obtained by fitting the decay of the Larmor revivals (the dashed curves in Fig.\,\ref{fig2}),
while the open squares represent the fit to the initial decay.
In both cases, the decay time increases with the number of pulses in the DD sequence.
If we take the values from the maxima of the Larmor revivals, they increase from $\approx 0.3$ ms for the Hahn echo
up to 2.2 ($\pm$0.2) ms for CPMG 128, which is the longest value ever reported for the NV center.

Apart from the increase of the decoherence time with the number of pulses, we find that also
the shape of the decay curve changes.
In Fig.\ref{fig3}, we quantify this by the exponent $n$, which changes from $\approx 3$ for the Hahn echo
to $\approx 1$ for CPMG 128.
This result indicates that different processes are responsible for the observed decoherence.
For relatively long pulse separations, as in the case of the Hahn echo,
the dominant environmental interaction is the hyperfine coupling to the nuclear spins.
The shortest relevant timescale for the nuclear spins is the Larmor precession, which has a period of
$\tau_L \approx 73 \, \mathrm{\mu s}$.
Since this evolution is coherent, the associated decay is refocused at specific times,
giving rise to the Larmor revivals \cite{Childress2006}.
On a slower timescale, the nuclear spins undergo mutual spin flips, which also contribute to the decoherence of the electron spin.

When the spacing between the pulses becomes short compared to the Larmor period,
the refocusing of the decoherence due to the nuclear spin bath becomes almost perfect.
This is demonstrated in Fig.\,\ref{fig3} by the convergence of the curves from the decay time of the envelope
with that of the initial decay.
Both time scales approach a limiting value of $\approx 2.2$ ms, which is only slightly shorter than the
longitudinal relaxation time for this spin, which is $T_1 \approx 4$ ms.
Apparently, the remaining decoherence processes causing this decay are no longer due to the interaction
with the nuclear spins, but arise from processes with significantly shorter correlation times,
which cannot be refocused by dynamical decoupling.
Possible candidates for these processes include interactions with paramagnetic centers, electric fields\cite{dolde2011}
or phononic processes.
This interpretation is compatible with the observation of a purely exponential decay,
which is expected if the correlation time of environment is shorter than interpulse delay of dynamical decoupling.

\subsection{Parallel and orthogonal initial states}

\begin{figure}
\includegraphics[width=\columnwidth]{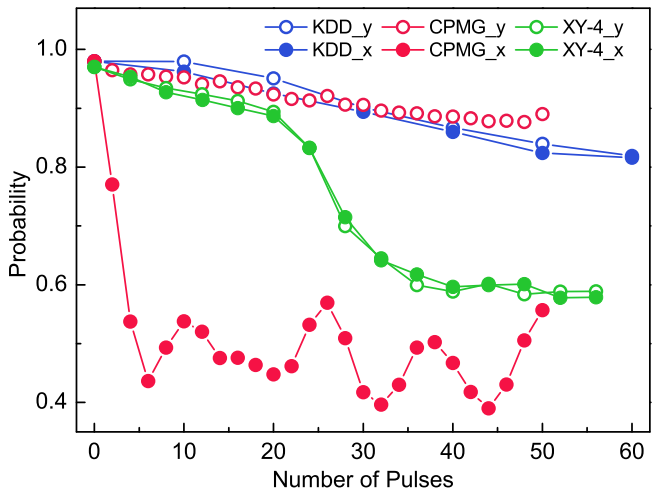}
\caption{(color online) Survival probabilities of two orthogonal initial states, $ |x \rangle$ (solid symbols)
and $| y \rangle$ (empty) during different sequences of equidistant refocusing pulses
for CPMG (red), KDD (blue), and XY-4 (green).
The time interval between the refocusing pulses was fixed to 0.8 $\mu s$.
}\label{fig4}
\end{figure}

The above experiments were performed with one specific initial condition,
for which the CPMG sequence is optimized.
However, in quantum information processing, the initial state is in general unknown,
and it is important to use a pulse
sequence that performs well for arbitrary initial conditions.
To evaluate the effectiveness for arbitrary initial conditions, we now consider the
effect of the DD sequence on two orthogonal initial conditions
$|x\rangle = (|0\rangle + |1\rangle)/\sqrt{2}$ and
$|y\rangle = (|0\rangle + i |1\rangle)/\sqrt{2}$.
Fig.~\ref{fig4} shows the decay of these states for the DD sequences CPMG, KDD, and XY-4
as a function of the number of refocusing pulses.
In the case of the CPMG sequence, we observe a strong asymmetry between the
two initial condition:
If the initial spin polarization is oriented along the rotation axis of the refocusing pulses,
the state preservation works extremely well.
If the initial state is perpendicular to the direction of the pulses, the first $\approx 10$ pulses
destroy the coherence almost completely.
This effect is expected if the rotation angle of the pulses deviates from the ideal value of $\pi$.\cite{alvarez2010}
Such a flip angle error has no effect on a state that is polarized along the rotation axis,
but for orthogonal orientations, the flip angle errors of the individual pulses accumulate
and cause an unwanted evolution of the system.

Such a strong dependence on the initial state of the system is not compatible with quantum information processing,
where the state is not known in general.
Universal DD sequences must perform well for arbitrary quantum states
and they have to be robust against pulse errors.
The earliest sequence that fulfills this requirement is the
XY-4 sequence \cite{Maudsley1986488,Viola2003}
and its derivatives that increase the robustness by combining different versions of the basic XY-4 cycle
into longer cycles with better compensation of pulse imperfections  \cite{Gullion1990479,Khodjasteh:2005it}.
A more recent sequence that shows even better performance is the KDD sequence \cite{souza2011},
which is based on an expansion scheme developed by Cho et al \cite{Cho1986}.
As shown in Fig.~\ref{fig4}, the performance of both sequences does not depend on the
initial state, within the experimental uncertainties.
The performance of the XY-4 sequence is comparable to that of CPMG\_y for $<20$ pulses but
deteriorates for longer sequences, while the KDD sequence matches the performance of CPMG\_y
for both initial conditions, indicating that this sequence largely compensates the flip angle errors
of the individual pulses.

\subsection{Arbitrary initial states}

\begin{figure}
\includegraphics[width=\columnwidth]{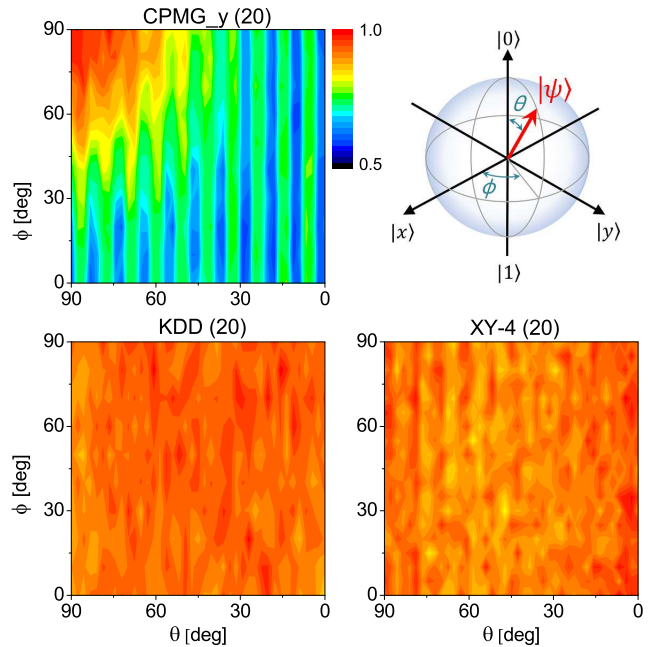}
\caption{(color online) Experimental survival probability for the sequences CPMG, KDD and XY-4
after 20 DD pulses as a function of the parameters $\theta$ and $\phi$
that define the initial state of eq. (\ref{rho_init}). Top left corner illustrates Bloch sphere representation of arbitrary state of NV center.
}\label{fig5}
\end{figure}

Fig.~\ref{fig5} extends this study to arbitrary initial conditions: for these experiments,
we prepared the initial states
\begin{equation}
|\psi_I\rangle = \cos(\frac{\theta}{2})|0\rangle + \sin(\frac{\theta}{2}) e^{i\phi} | 1 \rangle,
\label{rho_init}
\end{equation}
where the angles $\theta$ and $\phi$, which parametrize the state,
correspond to spherical coordinates on the Bloch sphere.
Fig.~\ref{fig5} shows the observed survival probability $|\langle \psi_F | \psi_I \rangle|^2$,
with $| \psi_F \rangle$ representing the state of the qubit after 20 refocusing pulses.
The three panels represent the survival probability as a function of $\theta$ and $\phi$
for the three DD sequences CPMG, XY-4 and KDD.
All three sequences reach almost perfect survival probabilities for $\theta = \phi = \pi/2$,
which corresponds to the $y$ initial condition where the spins are aligned with the rotation
axis of the CPMG pulses.
In the case of CPMG, the performance deteriorates when the initial condition deviates
significantly from this situation.
This behavior is an indication that the decay of the system fidelity is dominated
not by the environment, but rather by pulse imperfections \cite{alvarez2010,souza2011}.

\subsection{Effect of pulse imperfections}

\begin{figure}
\includegraphics[width=\columnwidth]{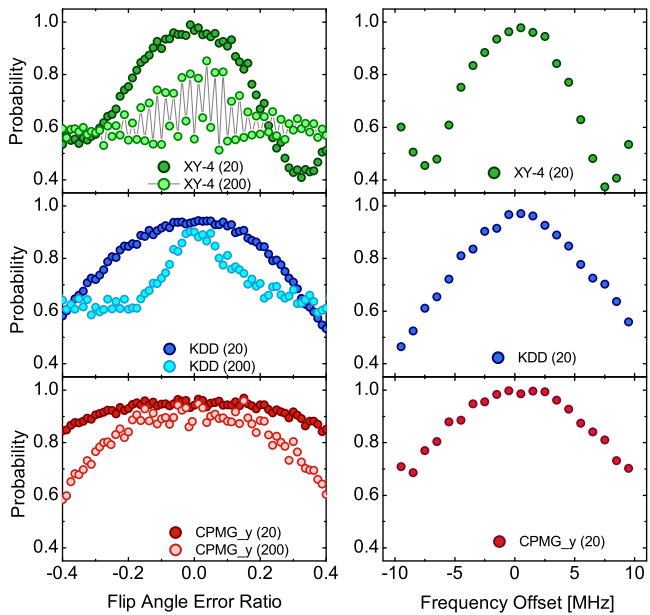}
\caption{(color online) Left column : Survival probability of the electron spin state as a function of
the fractional flip angle error.
The initial state was chosen along the rotation axis of the CPMG pulses.
The numbers in the parentheses represent number of refocusing pulses in DD sequence.
Right column : variation of survival probability as a function of frequency offset errors on pulses.
The numbers in the parentheses represent number of refocusing pulses in the DD sequence.}\label{fig6}
\end{figure}

Apart from finite pulse durations, the two dominant experimental imperfections that limit the performance
of experimental dynamical decoupling are flip angle errors and offset errors.
In Fig.~\ref{fig6}, we experimentally investigate the effect of these imperfections on the performance of
different dynamical decoupling sequences.
In the left hand columns, we measured the remaining spin polarization after 20 and 200 pulses as a function
of the actual flip angle of each pulse.
For 20 refocusing pulses and longitudinal initial conditions, flip angle errors
up to $\pm$ 10 $\%$ do not reduce the fidelity.
For higher errors, we find a bi-quadratic reduction (not shown here).
This  is consistent with theoretical considerations.\cite{borneman2010}
For the robust sequences, whose performance does not depend significantly
on the initial condition, KDD clearly outperforms XY-4.
This is particularly well visible for 200 pulses.
Interestingly, XY-4 shows strong oscillations as a function of the flip angle error.
This implies that XY-4 becomes very sensitive to flip angle errors if more than 100 refocusing pulses are used,
which limits its usability.
The right hand column shows the measured behavior as a function of the
offset error for 20 refocusing pulses.
For all three sequences, we observe significant reductions of the fidelity
if the offset exceeds $\approx 2$ MHz.
This is significantly less than the Rabi frequency (12.5 MHz) of the pulses used for these experiments
and comparable to inherent offsets like the hyperfine interaction with the nitrogen nuclear spin.

In order to better understand the processes that reduce the process fidelity in these cases,
we performed state tomography
after 20 refocusing pulses for the initial state along the $x$-axis ($\theta=\frac{\pi}{2}$, $\phi=0$).
From the state tomography, we could estimate a flip angle error of 1.6 $\%$,
which corresponds to 2.9 degree deviation from the targeted 180 degree rotation.
This value is quite consistent with our hardware limit on the width of time step.
The arbitrary wave form generator used for pulse generation has 0.5 ns step width (2 GS/s).
Given the duration of the refocusing pulse of 40 ns, the digital time resolution of the pulse
results in a flip angle error of 1.2 $\%$, which is not far from  value given above and does not
yet include effects due to amplifier nonlinearities or transient effects.
Frequency offset error was also observed, and this was - 4 $\%$ (0.5 MHz)
of the Rabi oscillation frequency (12.5 MHz).
This amount of estimation can be supported by the results of the frequency offset errors in Fig~\ref{fig6}.
The central positions of all the three curves are slightly shifted to positive frequency, and this deviation is around 0.5 MHz.
The origin of this frequency offset error is unclear to us,
because we tune the frequency of the microwave pulses to the center of the spectrum measured
by the Ramsey fringe experiment in Fig.~\ref{fig1} (d) as precisely as possible,
expecting that the off-resonance effect of the two outer hyperfine lines  cancel each other.
There is one way to confirm whether the off-resonance effect generates frequency offset errors on the pulses:
doing experiments at the excited state level anti-crossing point (514 G),
only one of the $^{14}$N nuclear spin states will be populated,\cite{Jacques2009}
and this will make it available to remove off-resonance effect completely.

\section{Conclusions}

Dynamical decoupling is an effective method for increasing the coherence time of quantum bits,
such as spin qubits in solids.
As long as the fluctuation of the noise sources is comparable to or slower than the system's coherence time,
DD can increase the survival time of the quantum information.
With CPMG sequences, we have shown that T$_2$ of a single NV center in natural abundance diamond
can be extended up to 2.2 ms by strongly suppressing the fluctuating fields from the $^{13}$C nuclear spin bath.
This is close to the T$_1$ limit, which is determined by different, more rapidly fluctuating processes.
We found that the recently introduced DD sequence KDD is remarkably robust and works for
arbitrary initial conditions.
It is much less susceptible to experimental uncertainties, such as flip-angle errors and
frequency offsets, than the simpler sequences CPMG and XY-4.
We expect that this robust DD sequence will be used for many other purposes requiring long coherence time,
such as ac-magnetometry or to increase the lifetime of multipartite systems, including entangled states.

\acknowledgments
We thank F. Jelezko for his generous loan of the sample and
Gonzalo A. \'Alvarez and Alexandre M. Souza for fruitful discussions and helpful advice.
This work was supported by the DFG through grant Su 192/27-1.

\bibliographystyle{apsrev}
\bibliography{RDD_NVC}
\end{document}